\documentclass[journal,onecolumn,draftcls,12pt,twoside]{IEEEtranTCOM}
\normalsize

\usepackage{amsfonts}
\usepackage{color}
\usepackage{amsmath}
\usepackage{xcolor}
\usepackage{graphicx}

\newlength\figurewidth
\definecolor{col1}{rgb}{0.0000,0.4470,0.7410}%
\definecolor{col2}{rgb}{0.8500,0.3250,0.0980}%
\definecolor{col3}{rgb}{0.9290,0.6940,0.1250}%
\definecolor{col4}{rgb}{0.4940,0.1840,0.5560}%
\definecolor{col5}{rgb}{0.4660,0.6740,0.1880}%
\definecolor{col6}{rgb}{0.3010,0.7450,0.9330}%
\catcode`,\active

\catcode`\,12

\title{High-Rate Regular APSK Constellations}
\author{Paul Ferrand,~\IEEEmembership{Member,~IEEE,} Marco Maso,~\IEEEmembership{Senior Member,~IEEE,} and Valerio Bioglio,~\IEEEmembership{Member,~IEEE}%
\thanks{P. Ferrand and V. Bioglio are with Huawei Technologies France S.A.S.U., 20 quai du Point du Jour, 92100 Boulogne Billancourt, France. M. Maso is with Nokia Bell Labs France, route de Villejust, 91620 Nozay, France.}}
\begin{document}
\maketitle
\begin{abstract}
The majority of modern communication systems adopts quadrature amplitude modulation (QAM) constellations as transmission schemes. 
Due to their square structure, however, QAM do not provide satisfying protection to phase noise effects as the number of constellation points grows, increasing at the same time their peak to average power ratio (PAPR). 
This requires an expensive power amplifier and oscillator at the transmitter to guarantee low distortion, complicating the adoption of dense transmission schemes in practical high-data rate systems.
In this paper, we construct a coded modulation scheme based on regular amplitude and phase shift keying (RAPSK) modulations.
We propose a novel multilevel coding (MLC) labeling for the constellation points separating amplitude and phase domains. 
We provide a novel multistage decoding (MSD) scheme allowing for a low-complexity log-likelihood ratio (LLR) calculation for soft-input decoding of component codes, along with a suitable rate design. 
Finally, we compare the proposed scheme with state-of-the-art QAM constellations and optimized constellations in the presence of phase noise. 
\end{abstract}
\IEEEpeerreviewmaketitle
\section{Introduction} \label{sec:intro}

\IEEEPARstart{T}{he adoption} of high data rate communication systems has become a necessity at many levels of modern networks, in order to be able to cope with the growing data and traffic demands of both end-users and operators.
In particular, researchers focused on the development of effective solutions to deploy high rate point-to-point solutions for both wireless fixed microwave wireless backhaul links and wired optical fiber links scenarios~\cite{Bohagen2007}. 
One approach is to increase the number of bits per symbol sent on a fixed bandwidth, since the signal-to-noise ratio (SNR) achieved on short distances potentially enables very high density constellations~\cite{Boiocchi2013}.
However, increasing the density of classical quadrature amplitude modulation (QAM) comes with several drawbacks in practical applications.
One major problem is that the peak to average power ratio (PAPR) of QAM constellations greatly increases with the number of constellation points.
The wide range of amplitude values of such constellations can result in saturation of the amplifier or alternatively encourage the use of large back-offs that impact power efficiency and subsequently SNR at the receiver~\cite{Pratt2006}. 
No satisfying solutions exist for single carrier systems, and most manufacturers use more expensive power amplifiers with the required dynamic range~\cite{Lim2009} or increase the backoff in their systems to compensate.

Furthermore, dense QAM constellations are very sensitive to phase noise, which rotates the received signal so that points farther away from the center are more affected than points closer to the center~\cite{Taggart2011}.
Unfortunately, classic QAM constellations do not permit to change the density of the constellation depending on the distance of the point from the origin.
In practical communication systems, the limited robustness to phase noise of QAM constellations can be compensated in part by using better oscillators or protecting the outer points through set partitioning~\cite{las_part}. 

On the other hand, circular constellations naturally adapt to this scenario; circularly symmetric 64-point constellation proposed in \cite{Ghosh1996} yields higher robustness to phase noise than QAM, while guaranteeing a high energy efficiency, however without targeting PAPR reduction.
Irregular Amplitude and Phase Shift Keying (APSK) constellations have been used in standards~\cite{dvbS2} and applications \cite{de2003coded}, typically employing Gray labeling to support bit-interleaved coded modulation (BICM) with a limited number of points.
While it is possible to design a coded modulation scheme for phase noise channels assuming that the phase noise is a general form of fading~\cite{Yadav2013}, using the specific structure of phase noise provides interesting guidelines into the design process.
Similar to the results of~\cite{Foschini1973}, recent works showed that optimized constellation designs based on mutual information in phase noise channels converges toward irregular APSK solutions \cite{Kayhan2014,Yang2013,Yan2013}.
However, since such solutions do not exhibit a predefined structure, their decoding complexity is still rather high, and they usually require careful and costly black-box optimization with respect to the system parameters.
A low density APSK constellation supporting Gray labeling has also been proposed in \cite{Liu2011}, which is suitable for BICM approaches~\cite{Xie2012} and called a product-APSK constellation.
The authors further show in~\cite{Yang2014} that product-APSK constellations can provide substantial gains over QAMs under white noise Gaussian channels in some SNR regimes.
However, the labeling of \cite{Xie2012} does not target high data rate applications for which the support for multilevel coding (MLC) approaches \cite{Imai1977} would be preferred.
Some specific irregular APSK with set partitioning have been developed in~\cite{Yoda2015}, however not allowing for straightforward generalizations to high density constellations.

\begin{figure}[t]
	\centering
	\includegraphics{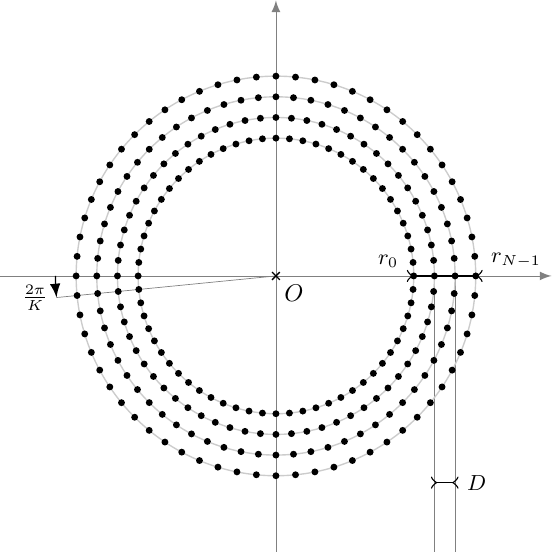}
	\caption{RAPSK constellation with $N = 4$ rings of $K = 64$ points.}
	\label{fig:cpp_example}
\end{figure}

In this paper, we design a dense Regular Amplitude and Phase Shift Keying (RAPSK) constellation for which the constellation points are arranged in concentric rings.
Similarly to~\cite{Liu2011,Xie2012,Yang2014}, each ring carries the same number of points, and points on different rings are aligned on semi-lines starting from the center; however, we further require adjacent rings to be equidistant. %, except the first ring which can be any distance away from the constellation center.
This construction provides an inherent robustness to phase shifts effect and can be parametrized for low PAPR.
The additional constraints with respect to product-APSK constellations leads us away from optimality with respect to channels affected only by white Gaussian noise~\cite{Yang2014}.
However, the main benefit of our proposal is that it allows for a labeling based on the MLC approach suitable for a soft demodulator whose complexity does not depend on the constellation size.
We describe the labeling and coding process, as well as a noise analysis under both white and phase noises.
We also provide a fast log-likelihood ratio (LLR) computation which has a fixed complexity and does not depend on the constellation size.
As a necessary component to maximize the performance of MLC approaches, we also describe a rate design for the bit channels in the MLC scheme.
The merit of this constellation and code design is finally assessed by means of Monte Carlo simulations, for which extended irregular repeat-accumulate low density parity check codes (eIRA LDPC)~\cite{yang2004} are used as component codes for the MLC scheme.

\section{Proposed Constellation Design}\label{SEC:SystemModel}

We construct the proposed RAPSK constellation as follows, and refer the reader to Fig.~\ref{fig:cpp_example}.
We initially take a segment lying on the abscissa of a two-dimensional Cartesian coordinates system having the two extremities located at distances $r_0$ an $r_{N-1}$. 
Next, we consider $N$ equidistant points over this segment, i.e., the first one at coordinates $(r_0, 0)$, the last one at $(r_{N-1}, 0)$ and $N-2$ points in between, each one at distance $D=(r_{N-1}-r_0)/(N-1)$ from its two closest neighbors. 
The other constellation points are obtained by rotating this original segment $K-1$ times.
By construction, the radius $r_n$ of the $(n+1)$-th ring and the angle $\theta_k$ of the $k$-th rotated version of the original segment are thus given by
\begin{equation}
	\label{eq:rn_thetak}
	r_n = r_0 + nD \qquad \theta_k = \frac{2\pi k}{K}.
\end{equation}
This design results in a constellation composed of $N$ concentric rings, each containing $K$ points which share a constant phase in groups of $N$, for a total of $M = NK$ constellation points. 
In the following, the number of constellation points $M$ is limited to powers of two to allow for MLC demodulation. 

The constellation is thus formed by $N$ concentric circles.
Given that there is an even number of points per ring, the power of the constellation, denoted as $P$, is equal to
\begin{align}
&P = \frac1N\sum_{n=0}^{N-1} r_n^2 = r_0^2 +  \frac{2Dr_0}N\sum_{n=0}^{N-1} n + \frac{D^2}N \sum_{n=0}^{N-1} n^2 \notag\\
&= r_0^2 + (N-1)r_0D  +  \frac{(N-1)(2N-1)}{6}D^2.
\label{eq:avg_power}
\end{align}
This controls the relationship between the parameters $N$, $D$, $r_0$ and the desired power, as a polynomial of degree 2 over either $r_0$ or $D$.
It is therefore possible to normalize the power through either of these parameters; we chose in the sequel to parametrize the constellation through the innermost radius $r_0$ and choose $D$ so that $P=1$ in \eqref{eq:avg_power}.
This results in an inter-ring distance that is a function of both $r_0$ and $N$ and may be written as
\begin{equation}
D = \frac{3r_0}{(2N-1)} \left( \sqrt{1+\frac{2(1-r_0^2)(2N-1)}{3r_0^2(N-1)}} -1 \right).\label{eq:opt_D}
\end{equation}

\subsection{Design rationale}

With Gaussian noise channels, the product-APSK constellation design is shown to be competitive in~\cite{Yang2014} in some SNR regimes.
However, the radii of the rings has to be chosen to mimic a Gaussian input distribution~\cite{Liu2011}.
Each channel of the product-APSK constellation can be decoded independently without incurring high performance loss in theory~\cite{Yang2014}, but building an efficient high-rate coding scheme is not straightforward.
Considering phase noise in addition to Gaussian noise, our constellation design aims at striking a good tradeoff between optimality and engineering flexibility.
Since all rings contain the same number of points, it is less sensitive to phase noise effects than QAM constellations.
Choosing $N$ and $K$ as powers of 2 along with an equidistant ring radius distribution enables advanced multilevel coding schemes, as detailed in the remainder of the article.
Finally, it is very well suited to high-gain power amplifiers, while packing rings away from the center reduces the PAPR without modifying the decoding process.
Furthermore, this constellation is easier to pre-equalize than equivalent QAMs~\cite{opt_book} by designing the circle radii to be equidistant \emph{after} equalization, and assigning a single phase rotation to each circle to compensate for the amplitude-dependent phase shifts.
All the equalization can thus be handled through only $N$ complex coefficients.
This engineering benefit is also shared by product-APSK constellations in general~\cite{Yang2014}.

\subsection{Constellation parameters analysis}
Overall, the proposed RAPSK constellations have more degrees of freedom in their design than QAM constellations.
The constellation parameters $N$, $K$ and $r_0$ can be freely chosen depending on the target performance and environmental parameters such as the average level of phase or white noise.
One could also use feedback on the detection performance to adapt the parameters, e.g. reducing the number of points per ring if the angular error rate is too high or decreasing the distance between the circles to reduce the PAPR if the radial error rate allows for it.
If we define the normalized ring distance as $\tilde D = D/r_0$, the PAPR of a RAPSK constellation can be calculated as 
\begin{equation}
PAPR = \frac{r_N^2}{P} = \frac{N\left[1 + (N-1)\tilde D\right]^2}{\sum_{n=0}^{N-1}\left[1 + n\tilde D\right]^2}.\label{eq:papr_equation}
\end{equation}
This relationship allows one to choose $\tilde D$ to reach a target PAPR: when the inter-ring distance $\tilde D$ is reduced towards 0, the PAPR tends to 1.
As $\tilde D$ increases towards infinity however, the terms in $\tilde D^2$ will dominate both the numerator and denonimator in \eqref{eq:papr_equation} and thus
\begin{equation}
	\label{eq:papr_asymptote}
	\lim_{\tilde D \to \infty} PAPR = \frac{N(N-1)^2}{\sum_{n=0}^{N-1} n^2} = \frac{6(N-1)}{2N-1}
\end{equation}
Finally, the choice of parameters is also linked to the desired code  rate, as described in section~\ref{sec:RD}.
It is possible to code at a lower rate to reduce the PAPR, or to accomodate for worse oscillators.

\subsection{Labeling method}
\label{sec:lab}
\begin{figure}[t!]
	\centering
		\includegraphics[width=0.6\columnwidth]{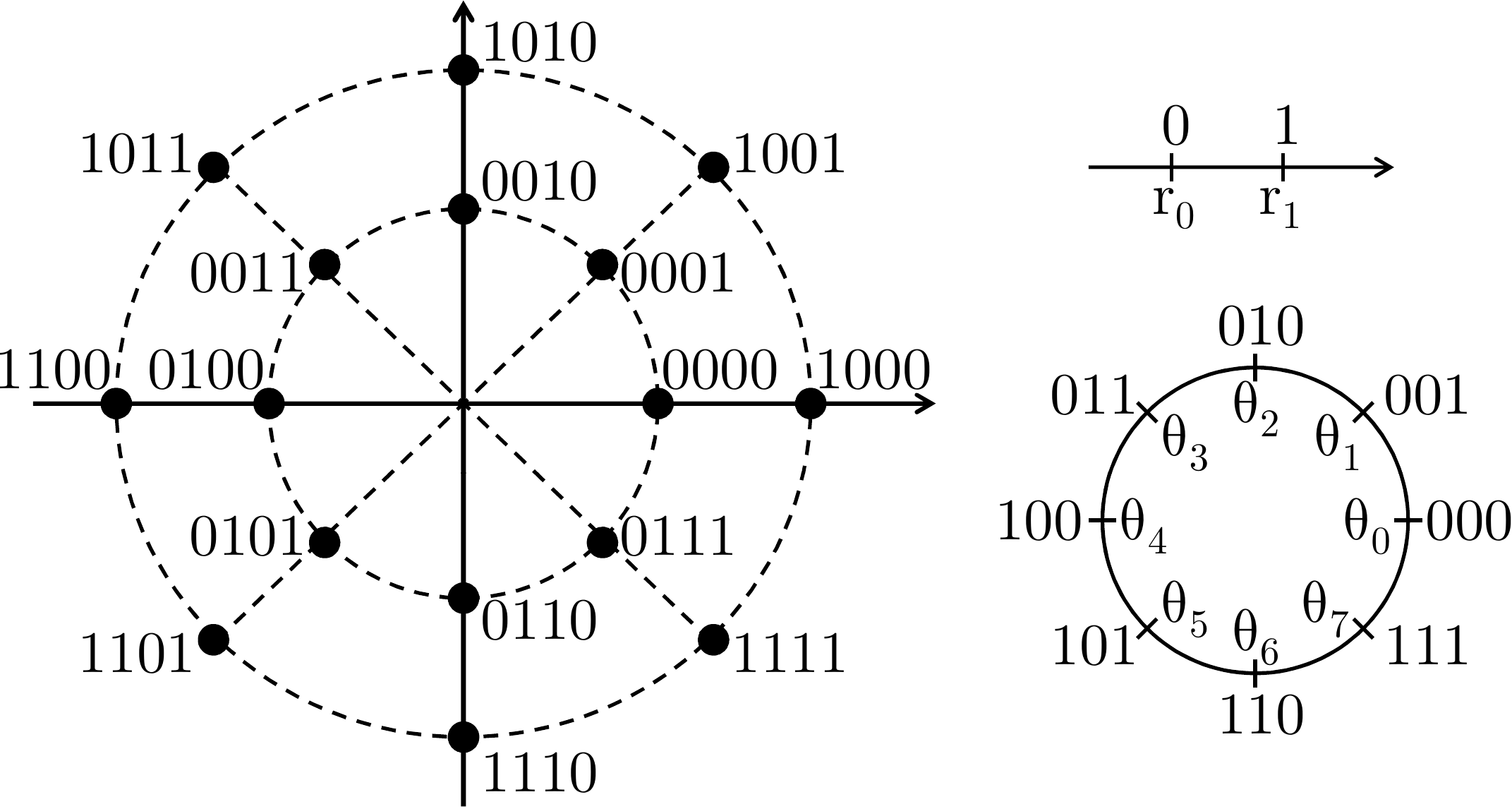} 
	\caption{Example of multilevel set partitioning and labeling for RAPSK constellations.}
	\label{fig:mlc}
\end{figure}
The $M=2^m$ constellation points are mapped into strings---or vectors---of $m$ bits. 
Even though classical Gray labeling can be extended to the presented RAPSK constellation~\cite{Liu2011,Xie2012}, we propose a different constellation labeling enabling a multilevel code (MLC) approach~\cite{MLC_intro}. 
Multilevel coded modulation permits to jointly optimize coding and modulation by protecting each of the $m$ bits with a different code $C_i$ at level $i$. 
The receiver performs multistage decoding (MSD), namely decoding each code individually starting from the first, while taking into account decisions of previously decoded levels. 

The proposed RAPSK constellation permits to further improve this framework by splitting the multilevel approach along the two different domains of the constellation. 
Labeling is hence decomposed in radial and angular domain such that the first $n$ bits identify the radius of the constellation point, while the last $k=m-n$ bits identify its angle. 
Labeling inside each domain is designed on the basis of the multi-level paradigm.  
In practice, using the definition in~\eqref{eq:rn_thetak}, a point $x = r_i e^{\jmath \theta_\ell}$ is represented by the string of bits $b_1 \dots b_m$, where $b_1 \dots b_n$ is straightforwardly given by the binary representation of the integer $i$ in $n$ digits and $b_{n+1} \dots b_m$ is given by the binary representation of the integer $\ell$ in $k$ digits. 
Fig.~\ref{fig:mlc} shows an example for 2 rings and 8 points on each ring, for which the constellation point given by $x = r_1 e^{\jmath \theta_2}$ corresponds to the string $1010$. 
We can readily see on Fig.~\ref{fig:mlc} that unlike Gray coding, adjacent points in the constellation do not differ by only a single bit; this is however of no consequence for MSD~\cite{MLC_intro}.

\section{Noise modeling} \label{sec:noise}

In this section, we study the effect of both phase and white noise on the points of the proposed RAPSK constellation. 
Noise will be decomposed in noise affecting the signal phase and noise affecting the signal magnitude.
This will be used to perform MSD at the receiver, as described in section~\ref{sec:demap}. 

Let $\mathcal C \subset \mathbb C$ be the finite set of the $M$ constellation points composing the RAPSK.
Assume that $x = \rho_x e^{\jmath \theta_x} \in \mathcal C$ is transmitted through a channel affected by both white noise and phase noise. 
Given a phase noise $\phi$ and the white noise $z$, the received signal $y \in \mathbb C$ is given by
\begin{equation}\label{eq:sys}
y = e^{\jmath \phi} x + z.
\end{equation}
The white noise component is distributed according to a circular complex Gaussian distribution $z \sim \mathcal{CN}(0,\sigma^2_z)$.
The phase noise is i.i.d. with a centered Von Mises distribution~\cite{Viterbi1963} $\phi \sim \mathcal{VM}(\kappa_{\phi})$ whose p.d.f. is 
\begin{equation}\label{eq:vonmises_dist}	
f_{VM}(\phi) = \frac{e^{\kappa_{\phi}\cos \phi}}{2 \pi I_0(\kappa_{\phi})}
\end{equation}
where $I_n(\kappa)$ is the modified Bessel function of first kind of order $n$~\cite{olver2010nist}.
The Von Mises distribution has bounded support in the angular domain, hence $\phi \in [0, 2\pi)$.

The combined effect of both phase and white noises on the received signal can be decomposed along the two polar components as
\begin{equation}\label{eq:sys_pol}
y = \rho_y e^{\jmath \theta_y} = (\rho_x + \rho_t)e^{\jmath (\theta_x + \theta_t)}, 
\end{equation}
where $\rho_t$ is the radial component of the composed noise and $\theta_t$ is the angular component of the composed noise. 
The radial noise $\rho_t$ mainly depends on the white noise $z$; the phase noise $\phi$ has a negligible effect on it compared to the white noise and will be not considered. 
On the other hand, the angular noise $\theta_t$ is impacted by both phase and white noises in different proportions. 
In particular, the relative weight of the white noise also changes with the distance from the center, i.e. depends on the ring of the transmitted point $x$. 
In this section, we show how to approximate $\rho_t$ and $\theta_t$ with equivalent Gaussian noise in order to simplify MSD at the receiver. 

\subsection{Radial component}

Let $y = y_R + \jmath y_I$ and $x = x_R + \jmath x_I$.
The joint distribution of the received signal $(y_R, y_I)$ is a non-central circular Gaussian with p.d.f.
\begin{equation}
f(y_R, y_I|x_R, x_I) = \frac{1}{2\pi\sigma^2_z} e^{-\frac{(y_R - x_R)^2}{2\sigma^2_z}}e^{-\frac{(y_I - x_I)^2}{2\sigma^2_z}}
\end{equation}
We drop the explicit dependence on $x$ for compactness in the following.
From the Jacobian of the polar components transformation we can derive the p.d.f. of the joint distribution of $(\rho_y, \theta_y)$ as
\begin{align}\label{eq:dist_joint}
f(\rho_y&, \theta_y)
= \frac{\rho_y}{2 \pi \sigma^2_z} e^{-\frac{(\rho_y \cos \theta_y - x_R)^2}{2\sigma^2_z}}e^{-\frac{(\rho_y \sin \theta_y - x_I)^2}{2\sigma^2_z}} \notag\\
&= \frac{\rho_y}{2 \pi \sigma^2_z} e^{-\frac{\rho_y^2 + x_R^2 + x_I^2}{2\sigma^2_z}} e^{\frac{\rho_y x_R \cos \theta_y + \rho_y x_I \sin \theta_y}{\sigma^2_z}}.
\end{align}
From \cite[Eq 3.937.3, p.496]{TISP} we know that 
\begin{equation}
	\label{eq:bessel_tisp}
	\int_0^{2 \pi}e^{\frac{\rho_y x_R \cos \theta_y + \rho_y x_I \sin \theta_y}{\sigma^2_z}} d\theta_y = 2\pi I_0\left( \frac{\rho_x \rho_y}{\sigma_z^2} \right)
\end{equation}
The marginal distribution of $\rho_y$ is then obtained integrating \eqref{eq:dist_joint} over $\theta_y$ using~\eqref{eq:bessel_tisp} and
\begin{equation}\label{eq:dist_r}
f(\rho_y) = \frac{\rho_y}{\sigma^2_z} I_0\left( \frac{\rho_x \rho_y}{\sigma_z^2} \right)e^{-\frac{\rho_y^2 + \rho_x^2}{2\sigma^2_z}}
\end{equation}
At high SNR, we would have $\rho_x \approx \rho_y$ and $\rho_x\rho_y/\sigma_z^2$ would also be large.
Using asymptotic results of Bessel functions~\cite[Ch.10.30]{olver2010nist} we have that
\begin{equation*}
f(\rho_y) \approx \frac{\rho_y e^{\frac{\rho_x \rho_y}{\sigma^2_z}} e^{-\frac{\rho_y^2 + \rho_x^2}{2 \sigma^2_z}}}{\sigma^2_z \sqrt{2 \pi \sigma^{-2}_z \rho_x \rho_y}} = \sqrt{\frac{\rho_y}{2 \sigma^2_z \pi \rho_x}} e^{-\frac{\left( \rho_y - \rho_x\right)^2}{2 \sigma^2_z}}
\end{equation*}
and the p.d.f. can thus be approximated with a Gaussian distribution $\mathcal N(\rho_x, \sigma^2_z)$; consequently we have that
\begin{equation}
\rho_y - \rho_x = \rho_t \sim \mathcal N(0, \sigma^2_z).
\end{equation}

\subsection{Angular component}
The effect of the white noise on the angular component of the composed noise depends on the magnitude of the transmitted symbol $x$. 
Assuming $\rho_x$ to be known e.g. through decoding, we can compute the conditional p.d.f. of the phase of $y$ using the relationship $f(\theta_y|\rho_y)f(\rho_y) = f(\theta_y, \rho_y)$.
We assume from here on that w.l.o.g. $x_R = -\rho_x$ and $x_I = 0$ and we let $\kappa_\rho = \rho_x \rho_y/\sigma^{2}_z$.
From \eqref{eq:dist_joint} and \eqref{eq:dist_r}, we thus have
\begin{equation}\label{eq:dist_theta}
f(\theta_y|\rho_y) = \frac{ e^{\kappa_\rho \cos \theta_y } }{2 \pi I_0( \kappa_\rho )}.
\end{equation}
which is exactly a Von Mises distribution as defined in~\eqref{eq:vonmises_dist}.
We know that in the limit of infinite $\kappa_\rho$, the Von Mises distribution tends to a Gaussian distribution of variance $\kappa_\rho^{-1}$.
However, we can obtain a finer asymptotic by applying a saddle-point approximation to the p.d.f. \eqref{eq:dist_theta}~\cite{Butler2007} to obtain a Gaussian p.d.f. with the same mean and a matching curvature rather than a matching variance.
This approximation is reasonable in the high SNR regime.
We know that the mean of $\theta_y|\rho_y$ is $\pi$, while the second derivative of its p.d.f. is
\begin{equation*}
\frac{d^2}{d\theta_y^2}f(\theta_y|\rho_y) = \frac{e^{-\kappa_\rho \cos \theta_y}}{2\pi I_0(\kappa_\rho)}\left( \kappa_\rho^2 \sin^2(\theta_y) - \kappa_\rho \cos \theta_y \right)
\end{equation*}
Evaluating this derivative at the mean $\theta_y = \pi$ gives
\begin{equation}
\left.\frac{d^2}{d\theta_y^2}f(\theta_y|\rho_y)\right|_{\theta_y = \pi} = \frac{\kappa_\rho e^{\kappa_\rho}}{2\pi I_0(\kappa_\rho)}
\end{equation}
On the other hand, we know that the second derivative of a Gaussian distribution $\mathcal N(\mu, \sigma^2)$ evaluated at the mean assumes the value $\left( \sigma \sqrt{2 \pi} \right)^{-1}$, from which we can obtain the variance of the approximate Gaussian distribution as 
\begin{equation}
\label{eq:sigma_w}
\sigma_w^2=\left(\frac{\kappa_\rho e^{\kappa_\rho}}{2\pi I_0(\kappa_\rho)} \right)^{-\frac23}.
\end{equation}
We can hence approximate the distribution of the phase offset due to the white noise as $\theta_t \sim \mathcal N(0, \sigma_w^2)$.
Note that for increasing SNR, $\sigma_w^2 \rightarrow \frac{\sigma^2_z}{\rho_x^2}$. 

Let's focus now on the effect of the phase noise $\phi$ on the angular component of the composed noise. 
At this stage, we can see that the distribution of $\phi$ has a close relationship with the distribution of the white noise in the angular domain in \eqref{eq:dist_theta}.
Using the previous approach, we can obtain a smoother saddle-point approximation where the variance of the approximate Gaussian is equal to
\begin{equation}
\label{eq:sigma_p}
\sigma_p^2 = \left(\frac{\kappa_\phi e^{\kappa_\phi}}{\sqrt{2\pi} I_0(\kappa_\phi)} \right)^{-\frac23}.
\end{equation}
Overall, since both noise contributions are independent, we can approximate the global noise of the phase component as a Gaussian noise $\theta_t \sim \mathcal N(0, \sigma_a^2)$, with equivalent variance
\begin{equation}
\label{eq:sigma_a}
\sigma_a^2 = \sigma_w^2 + \sigma_p^2.
\end{equation}

\section{Demodulation and code design}
\label{sec:code_design}

The MLC labeling proposed in Section~\ref{sec:lab} permits to separate the decoding of the radial and angular domains. 
In fact, the received symbol can be decomposed into the radial and the angular domains as \eqref{eq:sys_pol}, where $\rho_t \sim \mathcal{N}(0,\sigma_z^2)$ and $\theta_t \sim \mathcal{N}(0,\sigma_a^2)$ according to Section~\ref{sec:noise}. 
However, we showed in \eqref{eq:sigma_p} that $\sigma_a^2$ depends on the radius of the ring of the transmitted point: the radial domain has to be decoded first, in order to find the ring of the transmitted point. 
This information will be used to decode the angular domain. 
\subsection{MLC construction}

In the proposed scheme, $T$ transmitted symbols are encoded and decoded in one step. 
This corresponds to the transmission of $TM$ bits. 
A multilevel code is designed across the symbols to improve the transmission capability. 
More in detail, $M$ error correcting codes $C_i(T,H_i)$ are designed, i.e. each code outputs a bit sequence of length $T$ given an input sequence of $H_i$ bits. 
The rate design of those codes, i.e. how to choose their dimensions $H_i$, will be discussed in the Section~\ref{sec:RD}. 
Put together, these codes can transmit $H = \sum_{i=0}^{M-1} H_i$ information bits using $TM$ bits; the overall rate of the transmission is therefore $H/TM$.

Every time a string $U$ composed of $H$ information bits has to be transmitted, these bits are divided into $M$ bit strings $u_0,\dots,u_{M-1}$ composed by $H_0,\dots,H_{M-1}$ bits respectively. 
Each bit string $u_i$ is then encoded using the error correcting code $C_i(T,H_i)$, obtaining $M$ codewords $x_0,\dots,x_{M-1}$ of length $T$ bits each. 
The codewords are re-arranged as rows of a $M \times T$ binary matrix $X$. 
Finally, every column of $X$ is modulated according to the RAPSK as described in Section~\ref{sec:lab} and transmitted. 

\subsection{Computing LLRs along a domain}
\label{sec:llr}

According to the MLC framework~\cite{MLC_intro}, the error correcting codes $\{C_i\}$ are used sequentially to perform a hard demodulation of each level $i$ which is then used as a basis to decode the next level. 
All of the codes $\{C_i\}$ need soft-input LLRs in order to perform decoding. 
According to the MSD framework, the LLRs for decoding of code $C_i$ are calculated on the basis of the received signal and on the hard-output decisions on previously decoded codes. 
This LLRs calculation process is in general a complex task, whose computational complexity depends on the number of constellation points~\cite{univ_soft_demap}. 
In the following, we show that the proposed RAPSK heavily simplifies this task by permitting to calculate LLRs independently on the number of constellation points. 

In the following, we describe the LLR calculation along a single domain for the proposed constellation. 
We consider a general $2^Q$-ASK scheme transmitting a symbol $x$, that is received as $y = x + z$ where $z \sim \mathcal{N}(0,\sigma^2)$ is white noise.
According to \eqref{eq:sys_pol}, this model is valid for both the radial and the angular domain, and the proposed scheme can be used for soft demodulation of both components using different noise variances. 
We suppose that the transmitted symbol is in the form $x = \sum_{k=0}^{Q-1} b_k 2^{k}$, where $b_i = \{ 0,1 \}$ and $b = b_0 b_2 \dots b_{Q-1}$ is the binary expansion of $x$---so that there is a one-to-one mapping between the constellation set $\mathcal C$ and $\{ 0,1,2,\dots,2^Q-1\}$.

The LLRs are calculated bit-by-bit in a sequential order.
In practice, the LLR of a bit $b_i$ is calculated on the basis of the received signal $y$ and the previously decoded bits $b_0,\dots,b_{i-1}$.
If we call $x_i = \sum_{k=0}^{i} b_k 2^{k}$, we have that the LLR of bit $b_i$ can be calculated as
\begin{equation}\label{eq:LLR}
LLR_i = 
\log \left( \frac{\mathbb{P}(y_i | x_{i-1},b_i=0)}{\mathbb{P}(y_i | x_{i-1},b_i=1)} \right).
\end{equation}
Due to the system model, the a posteriori probability (APP) for bit $b_i$ can be calculated as
\begin{align}
\mathbb{P}(y_i &| x_{i-1},b_i) =\frac{1}{\sqrt{2 \pi} \sigma} \sum_{k=0}^{2^{Q-i}} e^{-\frac{(y-x_{i-1}-b_i2^{i-1}+2^{i}k)^2}{2 \sigma_i^2}} \notag\\
&=\frac{1}{ 2^{i-1} \sqrt{2 \pi} \sigma_i} \sum_{k=0}^{2^{Q-i}} e^{-\frac{(y_i-b_i+2k)^2}{2 \sigma_i^2}}  \notag\\
&\approx \frac{1}{ 2^{i-1} \sqrt{2 \pi} \sigma_i} \sum_{k = -\infty}^{+ \infty} e^{-\frac{(y_i-b_i+2k)^2}{2 \sigma_i^2}} \label{eq:P_y}
\end{align}
where
\begin{equation}
\label{eq:sigma_y}
\sigma_i = \frac{\sigma}{2^{i-1}} \qquad y_i = \frac{y-x_{i-1}}{2^{i-1}}. 
\end{equation}
The straightforward calculation of the LLRs depends on the number of points of the constellation.
However, through the last step of \eqref{eq:P_y} we can approximate the LLR computation using wrapped distributions to make it independent on the number of constellation points. 
A wrapped probability distribution is a continuous probability distribution defined on points on a unit circle. 
This will allow us to approximate \eqref{eq:LLR} with a compact expression. 

The wrapped normal distribution p.d.f can be described through the mean and the standard deviation of the unwrapped normal distribution $\mathcal{N}(0,\sigma^2)$ as 
\begin{equation}
f_{WN}(\theta | \mu,\sigma) = \frac{1}{\sqrt{2 \pi} \sigma} \sum_{k = -\infty}^{+ \infty} e^{-\frac{(\theta-\mu+2k\pi)^2}{2 \sigma^2}}
\end{equation}
for $\theta \in [ -\pi,\pi ]$.
As a consequence, we can rewrite \eqref{eq:P_y} as
\begin{align}
\mathbb{P}(y_i | b_i) &= \frac{1}{2^{i-1} \sqrt{2 \pi} \sigma_i} \sum_{k = -\infty}^{+ \infty} e^{-\frac{(y_i-b_i+2k)^2}{2 \sigma_i^2}} \notag\\
&= \frac{\pi}{2^{i-1} \sqrt{2 \pi} \sigma_i \pi} \sum_{k = -\infty}^{+ \infty} e^{-\frac{(y_i\pi-b_i\pi+2k\pi)^2}{2 \sigma_i^2 \pi^2}} \notag\\
&= \frac{\pi}{2^{i-1}} f_{WN} (y_i\pi | b_i\pi,\sigma_i \pi).
\end{align}
Expressing the probability \eqref{eq:P_y} in terms of wrapped normal distribution permits to approximate it using the Von Mises distribution, which has a more tractable expression~\eqref{eq:vonmises_dist}, knowing that $f_{WN}(\theta | \mu,\sigma) \approx f_{VM}(\theta | \mu,\kappa)$ where $\kappa = A^{-1} \left( e^{\frac{\sigma^2}{2}} \right)$ and $A(\kappa) = I_1(\kappa)/I_0(\kappa)$.
We can thus approximate \eqref{eq:P_y} further as
\begin{align}
\mathbb{P}(y_i | x_{i-1},b_i)\approx \frac{\pi}{2^{i-1}} f_{VM}(y_i\pi | b_i\pi,\kappa_i)
\label{eq:P_y_vs_VM}
\end{align}
with $\kappa_i = A^{-1} \left( e^{\frac{\sigma_i^2 \pi^2}{2}} \right)$. 
Finally, we can now rewrite \eqref{eq:LLR} as 
\begin{align}
LLR_i
&= \log \left( \frac{\frac{\pi e^{\kappa_i \cos(y_i\pi)}}{2^i \pi I_0(\kappa_i)}}{\frac{\pi e^{\kappa_i \cos(y_i\pi - \pi)}}{2^i \pi I_0(\kappa_i)}} \right) \notag\\
&= \kappa_i (\cos(y_i\pi) - \cos(y_i\pi - \pi)) \notag\\
&= 2 \kappa_i \cos(y_i\pi) \notag\\
&= 2 a(\sigma_i) \cos(y_i\pi) \label{eq:LLR_app}
\end{align}
where $a(t) = A^{-1} \left( e^{\frac{t^2 \pi^2}{2}} \right)$ is an auxiliary function that can be computed offline and tabulated to speed up the calculation.
Some results also exist to compute ratios of Bessel functions online~\cite{olver2010nist}.

This result can be used to simplify the LLRs calculation in a MSD decoder as follows. 
According to MSD framework, calculation of $LLR_i$, namely the LLR of bit $b_i$, is performed using \eqref{eq:LLR_app}. 
Parameters $y_i$ and $\sigma_i$ of this equation are calculated on the basis of received symbol $y$ and channel noise $\sigma$ as described in \eqref{eq:sigma_y}; $x_{i-1}$ is estimated on the basis of the previously decoded levels $b_0,\dots,b_{i-1}$ as  $x_{i-1} = \sum_{k=0}^{i-1} b_k 2^{k}$. 
Since the auxiliary function $a(\cdot)$ is tabulated offline and does not depend on the constellation size, the proposed method to calculate LLRs is independent of the number of points in the constellation. 
The same number of operation are required to demodulate each bit, which implies that the demodulation latency is constant in each level and the overall complexity is only dependent on the number of levels in the constellation. 

\subsection{Demapper under MLC Design}
\label{sec:demap}

Equipped with \eqref{eq:LLR_app}, now we can describe the retrieval of the $M \times T$ matrix $X$ in detail. 
Once the $T$ symbols are received, demapping begins. 
The first $n$ rows of $X$, belonging to the radial domain, are initially decoded. 
For every received symbol $y=\rho_y e^{j \theta_y}$, the LLR of the first bit $LLR_0$ is calculated using \eqref{eq:LLR_app} with $y_0 = \frac{\rho_y-r_0}{D}$ and $\sigma = \frac{\sigma_z}{D}$ as input: these correspond to the LLRs of the first row of $X$.
These $T$ LLRs are used by the soft-input/hard-output decoder of code $C_0$ to calculate the $H_0$ bits string $\hat{u}_0$, which represent the estimation of the input string $u_0$. 
This string is then re-encoded through $C_0$ to obtain the codeword $\hat{x}_0$, which will be used to calculate $y_1$ and $\sigma_1$ as in \eqref{eq:sigma_y}. 
This procedure of calculating the LLRs using previously decoded bit, decoding of a row and cancellation of the decoded bit for next level is repeated until all the bits of the radial domain are decoded, namely until the $n$-th row of $X$. 

The decoding of the angular domain proceeds in a similar way, however using a different noise value $\sigma$ for every symbol, calculated on the basis of the radius $r$ of the ring calculated in the first part of the demapping. 
More in detail, after the radial demodulation it is possible to estimate the magnitude $\rho_x$ of every received symbol, representing the ring radius of the transmitted constellation point. 
Angular demapping thus uses a different initial variance $\sigma$ required in \eqref{eq:sigma_y} for symbols belonging to different rings; this variance is calculated from \eqref{eq:sigma_w}, \eqref{eq:sigma_p} and \eqref{eq:sigma_a} as
\begin{align}
\sigma^2 = \frac{K^2}{4\pi^2}\sigma_a^2.
\end{align}
Recall in particular that $\sigma_a$ is a function of $\rho_x$, $\rho_y$, the white noise variance $\sigma_z^2$ and the phase noise parameter $\kappa_\phi$.
At the end of the process, $U$ is retrieved as $U = [\hat{u}_0 \hat{u}_2 \dots \hat{u}_{M-1}]$.

In this procedure, we use the approximations of section~\ref{sec:noise} as input to the LLR computation algorithms.
We expect these approximations to hold overall when the SNR is high enough to enable a small inter-ring distance $D$, and the innermost ring radius $r_0$ is large with respect to the inter ring distance.
For applications where high-density constellations are preferred, we expect the SNR to be large~\cite{Boiocchi2013}, and targeting lower PAPR values will ensure that both conditions will be met.

\subsection{MLC Rate Design}
\label{sec:RD}

An essential point in the design of a coded modulation scheme is the assignment of code rates to the component codes. 
Different approaches have been proposed in the literature~\cite{MLC_intro}, however in the following we design code rates according the capacity rule. 
According to this code construction method, code rate $H_i$ is chosen to be equal to the capacity of the equivalent binary symmetric channel (BSC) at level $i$. 
In order to design the rates of the codes $C_0,\dots,C_{M-1}$ for the MLC, the capacity of the equivalent BSC is calculated as follows. 

If $Q \rightarrow \infty$, the probability to commit an error in the decoding of a bit $c_i$ is given by
\begin{align}
p_i &= \sum_{k=-\infty}^{+\infty} \int_{2^{i-1} \left( 2k + \frac{1}{2} \right) }^{2^{i-1} \left( 2k + \frac{3}{2} \right)} f_{\mathcal N}(x|0,\sigma^2) dx\notag\\
&=\sum_{k=-\infty}^{+\infty} F_{\mathcal N}(2^i k + 3 \cdot 2^{i-2}|0,\sigma^2) \notag\\
&- \sum_{k=-\infty}^{+\infty} F_{\mathcal N}(2^i k + 2^{i-2}|0,\sigma^2) 
\end{align}
We can express this probability using the wrapped normal distribution as
\begin{align}
p_i 
&= \sum_{k=-\infty}^{+\infty} \int_{2^i k + 2^{i-2}}^{2^i k + 3 \cdot 2^{i-2}} f_{\mathcal N}(t|0,\sigma^2) dt\notag\\
&= \sum_{k=-\infty}^{+\infty} \int_{2^i k + 2^{i-2}}^{2^i k + 3 \cdot 2^{i-2}} \frac{1}{\sqrt{2 \pi} \sigma} e^{-\frac{t^2}{2 \sigma^2}} dt \notag\\
&= \int_{\frac{1}{2}}^{\frac{3}{2}} \sum_{k=-\infty}^{+\infty} \frac{1}{\sqrt{2 \pi} \sigma_i} e^{-\frac{(x+2k)^2}{2 \sigma_i^2}} dx
\end{align}
We obtain that the error probability of the equivalent BSC seen by the $i$-th bit is given by
\begin{align}
\label{eq:p_i}
p_i &= 2 \int_{\frac{\pi}{2}}^{\pi} f_{WN} (\psi | 0,\sigma_i \pi) d \psi \notag\\
&\approx 2 \int_{\frac{\pi}{2}}^{\pi} f_{VM} (\psi | 0,\kappa_i ) d \psi
\end{align}
with $\kappa_i = A^{-1}\big( e^{\frac{\sigma_i^2 \pi^2}{2}} \big)$, where $\sigma_i$ is calculated as explained in Section~\ref{sec:demap}.
For the radial domain, i.e. for $1 \leq i \leq n$, each code $C_i$ is designed with a rate $1-p_i$, hence with $H_i = (1-p_i) T$.
For the angular domain, i.e. for $n+1 \leq i \leq m$, the rate of the code $C_i$ can be calculated as the average of the rates of all the rings of the level. 
As a consequence, \eqref{eq:p_i} is used to calculate $p_i^{(r_j)}$ for all the rings $r_0,\dots,r_{N-1}$, and the code rate is given by
\begin{equation}
H_i = \frac{1}{N} \sum_{j=0}^{N-1} 1 - p_i^{(r_j)} .
\end{equation} 
Even though this procedure is asymptotically correct, it turns out to be too optimistic for finite block lengths. 
However, $H_i$ can be used as an upper bound on the achievable rate: fine tuning can be executed by simulations to optimize this parameter on the basis of the described theoretical results. 

\section{Performance analysis}

\begin{figure}[t!]
	\centering
	\includegraphics{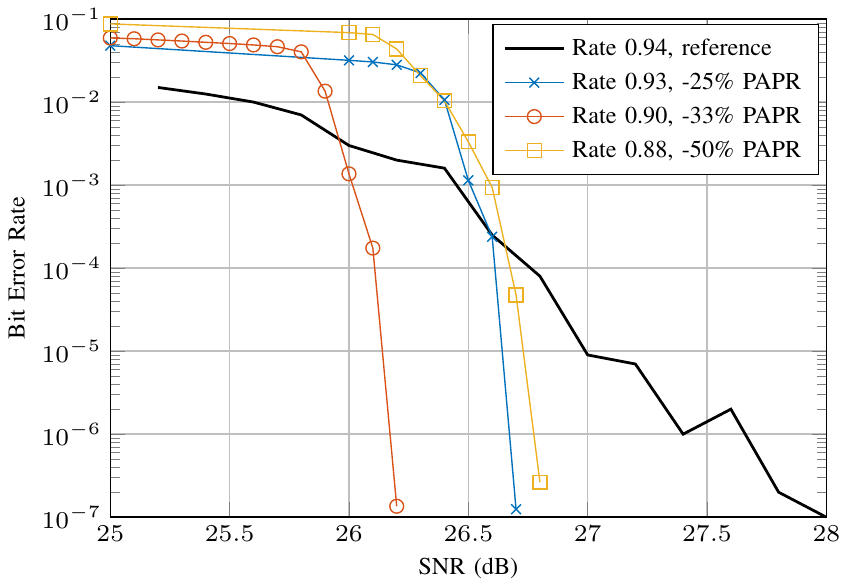}
	\caption{Performance of RAPSK and the optimized Kayhan-Montorsi constellation from \cite{Kayhan2014}.
	The constellation with rates 0.93 and 0.9 have parameters $K=32$, $M=8$ with $r_0=0.45$ and $r_0 = 0.55$ respectively.
	The constellation with rate 0.88 has parameters $K=64$, $M=4$ and $r_0=0.78$.
	}
	\label{fig:rapsk_kayhan}
\end{figure}

In this section, we present proof of concepts results and comparison of the proposed RAPSK constellation with respect to QAM constellations and constellation optimized for their robustness to phase noise~\cite{Kayhan2014}. 
All the presented results have been obtained by means of Monte Carlo simulations and additive white Gaussian noise (AWGN) channel with the additional phase noise as in~\eqref{eq:sys}, except for Fig.~\ref{fig:rapsk_kayhan} where the phase noise is generated as in~\cite{Kayhan2014} and thus not a memoryless process.
\begin{figure}[t!]
	\centering
    \includegraphics{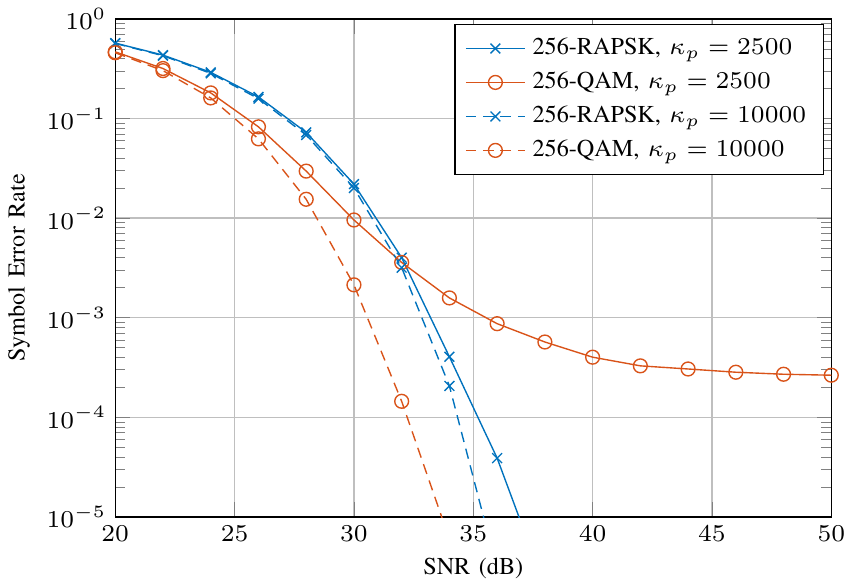}
	\caption{Comparison of a 256-RAPSK constellation with a 256-QAM constellation in presence of phase noise generated according to \eqref{eq:vonmises_dist} with different parameters $\kappa_p$. The RAPSK parameters are set as $N=8$, $K=32$ and $r_0 = 0.6$, which leads to a PAPR reduction of 32\% with respect to the QAM constellation.}
	\label{fig:uncoded_plot}
\end{figure}

In general, we see in simulations that QAM constellations, designed to be robust against white noise, are clearly better than RAPSK when phase noise is not the dominant source of errors.
On the other hand, correctly tuned regular APSK constellations will outperform QAM constellations when the phase noise induces a floor on the symbol error rate (SER) of the QAM-based transmission.
We analyze such a case in Fig.~\ref{fig:uncoded_plot}, where we compare a 256-QAM constellation with a 256-RAPSK constellations of the same size.
We focus on the symbol error rate (SER) without channel coding.
We apply two memoryless phase noise processes with parameters $\kappa_\phi = 2500$ and $\kappa_\phi = 10000$---a higher value for $\kappa_\phi$ means a lower variance for the phase noise process. 
In all cases the QAM constellations perform better at lower SNR in terms of SER.
When the phase noise variance is low, the QAM constellation outperforms the RAPSK constellation over the whole SNR range; the reduction in PAPR comes at the expense of performance in this case.
However, when phase noise becomes significant, the QAM constellation exhibits an error floor while the RAPSK constellation maintains a similar behavior to the one without phase noise.

\begin{figure}[b!]
	\centering
    \includegraphics{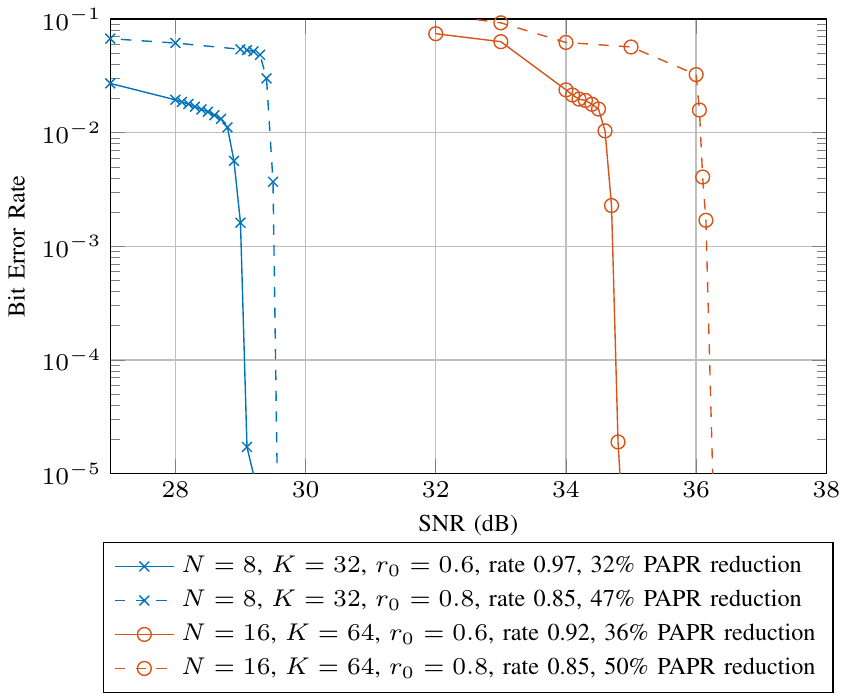}
	\caption{Comparison of RAPSK constellations with 256 and 1024 points using the parameters set in the legend and the MLC code described in section~\ref{sec:code_design}.} 
	\label{fig:rapsk_coded}
\end{figure}

Next, we tested the proposed MSD demapper for the RAPSK constellation and the associated rate design, to validate the approach and the flexibility of the RAPSK constellations.
The channel coding scheme used in our simulation is based on the eIRA LDPC~\cite{yang2004}.
The design of the component codes is built using $T=16200$ symbols per codeword, and a parameter $M$ that depends on the constellation size; for 256 points we have $M=8$ and for 1024 points we have $M=10$.
This scheme allows for rather efficient and fast simulations, however without offering a good flexibility in terms of code rates since the eIRA LDPC component codes only offer 10 rate levels with the highest one being limited to $8/9$.
As a consequence, the actual values of the adopted code rates are an approximation of the optimal ones provided by the rate design procedure of Sec.~\ref{sec:RD}, which is used as a guideline to select the rate of each component code.
Future channel code optimizations could certainly bring non-negligible gains in this regard.
We consider two constellation sizes with 256 and 1024 points respectively, under a strong phase noise profile with $\kappa_\phi = 1600$.
Under such a phase noise, QAM constellations will be severely distorted, and a custom set partitioning will be needed to single out the problematic points in the constellation.
On the other hand, the constellation and code design for RAPSK is straightforward.
Microwave links are usually held to stringent standards in terms of BER with target values as low as $10^{-10}$~\cite{Kizer2013}, which usually requires the channel code to have a sharp waterfall threshold~\cite{el2001analyzing}.
Applying the procedure of Sec.~\ref{sec:RD} leads to plunging BER curves, as can be seen in Fig.~\ref{fig:rapsk_coded}, which validates the proposed rate design and overall coding scheme.
In order to test the flexibility of the RAPSK design, we vary the innermost ring distance $r_0$ and the average code rate to trade off performance with the PAPR reduction, understood here with respect to a QAM constellation of the same size.
Using this approach, it is thus possible to tweak the constellation parameters to match target PAPR constraints and average oscillator phase noise, possibly at the expense of code rate and BER performance.

Fig.~\ref{fig:rapsk_kayhan} shows the performance of different RAPSK constellations versus the optimized Kayhan-Montorsi constellation of~\cite{Kayhan2014}.
The reference curve is taken from the original paper.
To match the reference simulation settings, we consider a realistic phase noise profile which corresponds to the one given in \cite{Kayhan2014} with an inflexion point set at -83 dBc.
The design of Kayhan and Montorsi optimizes both the constellation and the labeling of the constellation.
They then apply a state-of-the-art error correcting code with rate of $15/16 \approx 0.94$ on the optimized constellation.
Matching the code rate approximately---on the crossed curve of Fig.~\ref{fig:rapsk_kayhan}---we see that the RAPSK constellation design has a steeper BER curve; it shows relatively worse performance at lower SNR but is competitive in the regime of interest, and decreases rapidly as the SNR increases, as required for high-rate applications~\cite{Kizer2013}.
This constellation design also provides a reduced PAPR of 25\%.
We can further lower the PAPR and still remain competitive at this regime as evidence by the squared curve of Fig.~\ref{fig:rapsk_kayhan} using a lower code rate of about $7/8$ for a reduction in PAPR of around 50\%.
This reduction stems from a change from $K=32$ to a denser $K=64$ points per ring, which is sustainable with an appropriate coding rate in the angular domain.
Finally we can also aim for a lower waterfall SNR by trading off some of the code rate down to about $9/10$, as shown on the circled curve of Fig.~\ref{fig:rapsk_kayhan}.

\section{Conclusion}

In this paper, we proposed a coded modulation scheme based on regular APSK constellations.
By adding a limited set of constraints on the constellation and an appropriate labeling, we designed a multi-level channel code that is both low complexity and has competitive performance with respect to other state-of-the-art coded modulations.
The RAPSK constellations are flexible in their design and have several desirable characteristics from an engineering point of view, most notably with respect to PAPR, robustness to phase noise, and pre-equalization.
Moreover, they can be constructed in fully scalable way up to an arbitrarily high number of points, and set-partitionings of the points are naturally yielded by the structure itself. 
This allows to obtain constellations whose number of points is any power of 2 without substantial changes in the set-partitioning, coding and decoding procedure, contrary to QAM constellations for which the number of points is typically chosen as an even power of 2 to simplify detection and equalization procedures. 

In followup works, we plan to analyze further the optimal choice of parameters for RAPSK constellations with respect to key environmental parameters like the target PAPR and phase noise profile, using the theoretical analysis of the component BSC described in section~\ref{sec:RD}.
More exhaustive tests with more advanced channel coding is also warranted in order to understand the benefits and limits of this coded modulation scheme.
Finally, we plan to assess the robustness of the proposed design to fading effects in the channel, and thus evaluate the potential of this approach in a different scenario.

\bibliographystyle{IEEEtran}
\bibliography{IEEEabrv,tcom_final}

\end{document}